\documentclass[10pt,twoside]{article}
\usepackage{epsfig}
\def\deg       {$^o$}

\newcommand{\lsim}{\ \raise -2.truept\hbox{\rlap{\hbox{$\sim$}}\raise5.truept\hbox{$<$}\ }}
\newcommand{\gsim}{\ \raise -2.truept\hbox{\rlap{\hbox{$\sim$}}\raise5.truept\hbox{$>$}\ }}
\def\etal{{et\,al. }}
\def\degs{\ifmmode ^{\circ}\else$^{\circ}$\fi}
\def\amin{\ifmmode ^{\prime}\else$^{\prime}$\fi}
\def\asec{\ifmmode ^{\prime\prime}\else$^{\prime\prime}$\fi}

\newbox\grsign \setbox\grsign=\hbox{$>$}
\newdimen\grdimen \grdimen=\ht\grsign
\newbox\laxbox \newbox\gaxbox
\setbox\gaxbox=\hbox{\raise.5ex\hbox{$>$}\llap
     {\lower.5ex\hbox{$\sim$}}}\ht1=\grdimen\dp1=0pt
\setbox\laxbox=\hbox{\raise.5ex\hbox{$<$}\llap
     {\lower.5ex\hbox{$\sim$}}}\ht2=\grdimen\dp2=0pt
\def\gax{$\mathrel{\copy\gaxbox}$}
\def\lax{$\mathrel{\copy\laxbox}$}

\begin{document}

\thispagestyle{empty}

 \markboth{\it  ... \hfill (Greiner et al.: Nuclear Resonances)}{\it  ... \hfill (PI: J. Greiner) \hfill}

\centerline{\huge\bf Nuclear Resonances:}
\bigskip
\centerline{\LARGE\bf The quest for large column densities and a new tool}

\vspace{1.5cm}

\centerline{\Large\bf ``White Paper'' in support of Astro2010:}
\centerline{\Large\bf The Astronomy and Astrophysics Decadal Survey}

\vspace{4.0cm}
{\large
\begin{center}

coordinated by  J. Greiner\\
\smallskip
 Max-Planck-Institut f\"ur extraterrestrische Physik, Garching

\vspace{1.0cm}

with contributions from:
\end{center}

\medskip

\hspace{2.2cm}\parbox[c]{12.8cm}{\large
\noindent
S.E. Boggs -- SSL Berkeley (USA) \\
G. DiCocco -- IASF Bologna (Italy) \\
K.T. Freese -- University of Michigan (USA) \\
N. Gehrels -- Goddard Space Flight Center (USA) \\
D.H. Hartmann -- Clemson University (USA) \\
A. Iyudin -- Moscow State University (Russia) \\
G. Kanbach -- MPE Garching (Germany) \\
A.A. Zdziarski -- Copernicus Center Warszawa (Poland) 
}

}

\vspace{4.cm}

\begin{center}
{\Large\bf Abstract:}
\end{center}
{\large Nuclear physics offers us a powerful tool: using nuclear resonance 
absorption lines to infer the physical conditions in astrophysical
settings which are otherwise difficult to deduce. Present-day
technology provides an increase in sensitivity over previous gamma-ray
missions large enough to utilize this tool for the first time. 
The most exciting promise is to measure gamma-ray bursts from the
first star(s) at redshifts 20--60, but also active galactic nuclei
are promising targets.}

\newpage

\normalsize

\centerline{\Large\bf Nucleonic Absorption-Line Spectroscopy}

\vspace{0.5cm}

We propose to add the utilization of gamma-ray absorption line spectroscopy
to the astronomical toolbox. Nuclear level transitions carry their
own specific information, which can complement studies in particular
of violent and embedded objects such as GRBs and nuclei of active galaxies.

\bigskip

\noindent
{\Large\bf I. The physical effect}

\vspace{0.3cm}

\noindent
{\large\bf I.1 Nucleonic cross sections}

\medskip

The detection of (resonant) absorption lines is the most frequently used
tool  for studying matter towards an astrophysical source 
at low and high redshifts
as illuminated by distant background sources such as quasars.
The depth and shape of these absorption lines tell us about the
physical conditions of gas located between the source and the observer. 
These are
used to derive densities, velocities and metallicities, in order to 
constrain and unravel
cosmological structure and evolution.

Similar to X-ray and optical absorption lines which are due to transitions
between electronic levels, resonant absorption processes in atomic nuclei
exist which leave characteristic absorption lines in the $\gamma$-ray range.
The most prominent and astrophysically relevant are the nuclear excitation and
Pygmy resonances (element-specific narrow lines between 5--9 MeV),
the Giant Dipole resonance  (GDR; proton versus neutron fluid oscillations;
$\sim$ 25 MeV; two nucleons and more)
and the Delta-resonance (individual-nucleon excitations,
325 MeV; all nucleons, including H!).
The following is a short description of each of these astrophysically
relevant resonances - a more detailed description can be found in
Iyudin et al. (2005, A\&A 436, 763).

{\bf Delta Resonance:}
At photon energies exceeding the threshold for pion production, the
total absorption cross section of a photon interacting with an individual
nucleon or with a nucleus shows a remarkably universal feature, a resonance
that corresponds to the isovector magnetic dipole transition that connects
the nucleon and the ${\Delta }$(1232) isobar.
The position of the ${\Delta }$-resonance in the photon absorption cross
section is $\approx$305 MeV for protons and
327$\pm$5 MeV for nuclei from helium up to uranium;
the width is somewhat larger for nuclei as compared to that of protons.

{\bf Giant Dipole Resonance:}
First observed in 1947 in photonuclear reactions,
the GDR is a collective oscillation of all protons against all neutrons in a
nucleus, and as such does not occur for hydrogen. All other elements
contribute, and for A$>$4 the maximum of the cross section is in the 20-30 MeV
range. For a solar abundance medium, Helium provides the
largest contribution.

{\bf Pygmy Resonances:}
Resonance-like absorption below the photoproduction threshold can be produced
either via photon absorption to the excitation level of the nuclei
or via the photon capture into the so-called "Pygmy'' dipole resonance.
The majority of the abundant isotopes in the interstellar matter have the
ground state with a zero spin value and a positive parity; e.g.,
$^4$He, $^{12}$C, $^{16}$O, $^{24}$Mg, $^{28}$Si, and $^{56}$Fe,
producing a cross-section maximum around 7 MeV.
Single resonances are narrow, but in any realistic observing condition
many elements with cross-section maxima at slightly different energies
overlap - so this is the most challenging resonance for observational
astronomy.

{\bf Nuclear Level Transitions:}
These are the more conventional analogue to atomic line transitions,
if the nuclear shell model is adopted. They cover the energy range
between about 0.5 and 8~MeV, one prominent example being the 4.430~MeV
line of $^{12}$C excitation.

\begin{figure}[ht]
\includegraphics[width=.45\textwidth, angle=0]{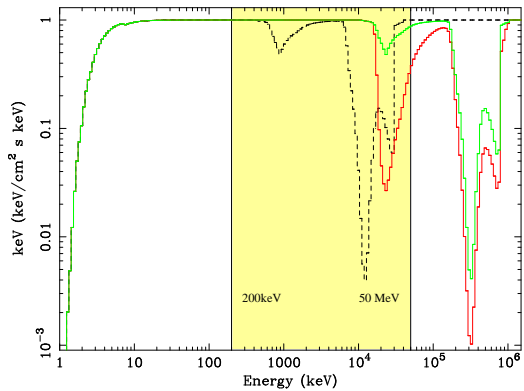}
\hfill\parbox[t]{7.2cm}{\vspace*{-5.5cm}\caption[Resonance absorption at 
different Z and z]{A flat $\nu$F$_\nu$ spectrum with N$_H$ = 10$^{28}$ cm$^{-2}$ 
resonance
absorption lines for two different redshifts (black: z=25; color: z=0)
and different metallicities: Z=0.1 (green) and Z=1 (red) solar metallicity.
Note the obvious difference in the relative strengths of the absorptions.
The solid vertical lines bracket the energy band
from 200 keV to 50 MeV which would be ideal to measure resonances
in the high-redshift Universe. Some galactic foreground absorption has been
been included (curvature of the green line at the very left).
\label{reso}}}
\vspace{-0.4cm}
\end{figure}

One extremely important property is that all the above
 resonant absorptions only depend on the presence of the
nucleonic species, and not on ionization state nor temperature.
This is different to electronic transitions in atomic shells. 
A draw-back, at least in the
past, is that large column densities are required to detect these
absorption features, due to relatively-low overall cross sections
combined with rather poor instrumental opportunities: 
past detector technology would have required
of order 10$^{28}$ cm$^{-2}$, while present-day technology (when flown
in 5--10 years) is able to detect column densities of 10$^{25}$ cm$^{-2}$.
Since in general the continuum spectra of $\gamma$-ray sources such as 
GRBs or blazars  
are otherwise featureless, these resonances imprint well-defined
spectral features which provide information which is unaccessible
otherwise.

\bigskip

\noindent
{\large\bf I.2 Cosmic column densities and scattering: What is the limit?}

\medskip

When studying absorption in a given source with either UV or X-ray
observations, one finds that the optical depth derived from the UV line
is always $\sim$50 times smaller than the X-ray derived depth
(e.g. Arav et al. 2002, 2003 in application to quasars).
This is a consequence of the wavelength difference between absorption lines
in UV and X-rays for the same ion, which means that X-rays are
sensitive to a much higher column density than the UV, and can be used
to provide a saturation test for the UV absorptions. Observations at soft
X-rays in
the early 90ies have expanded the maximum known source-intrinsic absorption
values from 10$^{19}$ cm$^{-2}$ to 10$^{22}$ cm$^{-2}$\footnote{Hydrogen 
absorption N$_H$ in units of 10$^{21}$ cm$^{-2}$ scales roughly with 
extinction A$_V$; the local ISM density times the distance to the
Galactic Center corresponds to A$_V \approx$ 25 mag, or 10$^6$ in flux 
reduction.}. 
Similarly, the advent
of sensitive observations at harder X-rays (first up to 10 keV,
later at 20--40 keV) have shifted the maximum known absorption to
5--10$\times$10$^{24}$ cm$^{-2}$, first
for NGC 1068 (e.g. Matt et al. 1997, A\&A, 325, L13), recently
for many sources as seen with Swift/BAT.
Along the same line of arguments, $\gamma $-ray absorption will probe
even higher column densities, which thus can be used to critically review 
saturation in X-ray absorption lines.

This is completely new territory, but with the
great promise to
1) probe even higher column densities than possible in the past,
  thus leading to a better understanding of source geometries
  and conditions in various source classes, especially the ones
  emitting high-energy radiation and being deeply embedded,
  which are probably of prominent cosmological relevance;
2) measure redshifts directly from the gamma-ray spectrum,
  i.e. without the need for optical/NIR follow-up work for their 
 identification!
The most promising sources to observe this effect are, fortunately,
the brightest known $\gamma$-ray sources: active galactic nuclei (AGN)
in outburst (particularly
blazars), and gamma-ray bursts (GRBs).

Is the resulting absorption detectable? 
This question has three aspects: 
(1) What are required instrumental sensitivity and
source statistics? 
(2) What are the constraints from candidate sources we know? 
(see separate Sections below), and 
(3) Could different physical effects destroy
the obvious absorption-line features? We address the latter now.
If the density is too high,
multiple-scattering of higher energy photons could partly
fill the energy window of a resonance, thus smearing the absorption trough.
This may happen via Compton scattering or
 the cascading of high-energy photons.
While the total pair production or Compton scattering
cross sections are about a factor of 30-40 larger than the
peak cross section of the Giant Dipole or Delta Resonance,
the jet geometry in both, GRBs as well as blazars, over-compensates
for this statistical measure: it is the differential cross section
which matters.
At the Dipole Resonance energy,
the Compton-scattered photon beam has a full-width-half-maximum
of 16\degs, or 0.5 sr. For a 1\degs\ opening angle of the jet,
the resulting GDR absorption is a factor $\sim$12 more efficient
than the Compton re-scattering of higher energy continuum photons
into the beam. In addition, Compton scattering changes the energy
of the scattered photon by arbitrary large values, much greater
than the width of the nuclear resonance - this adds another factor
of E/$\Delta$E (\gax 3 for the GDR) in favor of the nuclear resonance
absorption. For pair production, only the latter effect comes into play.
Thus, as the bottom line, all environments with a density smaller than about
10$^5$  cm$^{-3}$ will not self-destroy the nuclear absorption feature 
by refilling of the lines due to scattering.
Even in situations with still higher density environments, 
there are two possibilities which further alleviate the problem
of re-filling:
(i) the transverse dimension of the absorber is less
than $\sim$1.5 attenuation lengths at the energy of the highest
attenuation value (Varier \etal\ 1986) or (ii) the absorber consists
of many clumps (clouds) of matter.

\bigskip
\noindent
{\Large\bf II. Gamma-Ray Bursts}

\vspace{0.3cm}

GRB afterglows are bright enough to serve
as pathfinders to the very early universe.
 Since long-duration GRBs are related to the death of massive
stars, it is likely that high-$z$ GRBs exist. Theoretical
predictions range between few up to 50\% of all GRBs being at $z>5$
(Lamb \& Reichart 2001, Bromm \& Loeb 2002),
and stellar evolution models suggest that 50\% of all GRBs occur
at $z>4$ (Yoon \etal\ 2006).
The polarisation data of the Wilkinson
Microwave Anisotropy Probe (WMAP) indicate
a high electron scattering optical depth, hinting
that the first stars formed
in the range 20\lax z \lax 60
(Kogut \etal\ 2003, Bromm \& Loeb 2006, Naoz \& Bromberg 2007).
Measuring GRB spectra with sufficient sensitivity in the $\gamma$-ray
range, the detection of resonance absorption by matter near the GRB
will allow us to determine redshifts up to 100, and thus
measure the death of these first stars.

Is there enough matter along the sight lines to GRBs?
Apart from galactic foreground extinction, relatively
little intrinsic extinction has been found in the afterglow
spectral energy distributions
of GRBs, both at X-rays and at optical/NIR wavelengths.
A recent combined analysis of Swift XRT and UVOT
data shows that the
absorbers associated with the GRB host galaxy have column densities
(assuming solar abundances) ranging from (1-8)$\times$10$^{21}$ cm$^{-2}$
(Schady \etal\ 2007).
There is evidence, both theoretical as well as observational,
that there is a substantial amount of matter along the line of sights
to GRBs. This applies to the local GRB surroundings as well as to
the larger environment of the host galaxy in which the GRB explodes.
Temporally variable optical absorptions lines of fine-structure
transitions indicate that (i) all material at distances within
a few kpc is ionized, most likely by the strong UV photon
flux accompanied with the emission front of the GRB,
and (ii) beyond this ionized region the absorbing column is still at a 
level of 10$^{21}$ cm$^{-2}$  (Vreeswijk \etal\ 2007, A\&A 468, 83).
The present-day measurement capabilities in the optical/NIR as well
as X-rays are not adequate to determine the density of local matter
around GRBs.
However, at $\gamma$-rays this matter will be measurable
through nuclear resonance absorption even though this matter is
simultaneously being ionized: the GRB gamma-ray radiation
has to pass through it - and it will suffer resonance absorption independent
on whether this material is ionized or not.

\begin{figure}[ht]
\includegraphics[width=.35\textwidth]{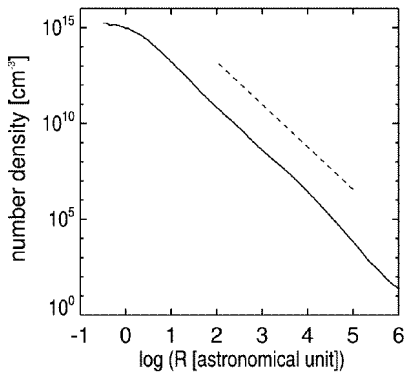}
\hfill\parbox[t]{7.5cm}{\vspace*{-3.7cm}\caption[Radial density at z=19]{Radial
 density around a GRB
progenitor at a redshift z=19. The density profile is close to
a power law $\propto$R$^{-2.2}$ (dashed line).
It contains about 10$^{28}$ cm$^{-2}$
column density within the inner 1-2 AU, and further
10$^{28}$ cm$^{-2}$ in each shell from 2-10, 10-100, 100-1000 AU!
(From Yoshida \etal\ 2006, ApJ 652, 6)
\label{GRBdens}}}
\vspace{-0.4cm}
\end{figure}

A variety of theoretical simulations of
GRB progenitors have been made (e.g.
Bate \& Bonnell 2003, 
Yoshida \etal\ 2006, ApJ 652, 6; 
Abel \etal\ 2007, ApJ 659, L87;
Gao \etal\ 2007, MN 378, 449),
pertaining to
the formation of the first stars, the fragmentation rate,
and density structure around the first stars.
The first stars are thought to form inside halos of mass
10$^5$...10$^6 M_{\odot}$ at redshifts 10-60.
It is generally accepted that most of the 10$^5$...10$^6 M_{\odot}$ halo mass
remains in the surroundings of the forming proto-star,
with about the original dimensions of the proto-cloud.
The resulting mass of the star as well as the density
structure are difficult to predict because they depend on the
collapse conditions (merger or not, strength of winds, etc).
However, it is important to realize that some simulations in fact predict
column densities of up to 10$^{29}$ cm$^{-2}$ around the first stars
(Yoshida \etal\ 2006,  Spolyar \etal\ 2008;  see also Fig. \ref{GRBdens}).
These simulations have been done irrespective
of nuclear resonance absorption.
It remains to be demonstrated (preferentially through observations) whether
the conditions in these simulations are realized.
Yet, the existence of what one ``normally''
would refer to as ``unbelievably high'' column densities is plausible -- note
that even pristine and fully-ionized hydrogen imprints resonant absorption!
GRBs are the best (and possibly only) tool to measure such conditions.

But even for low-redshift GRBs, high column densities are not
really excluded. In the standard picture, GRBs form in star-forming
regions. Furthermore, the distribution of GRB positions relative to
the centers of their host galaxies does not show large offsets.
Thus, GRB occur in or near the densest gas/dust regions in their hosts. From 
our own Galaxy we know that the clouds near the Galactic Center
are different from those in the Galactic disk: the incidence of dense
clouds is higher, with densities over 10$^4$ cm$^{-3}$
(Tsuboi et al. 1999, ApJS 120, 1). In particular, the
Midcourse Space Experiment (MSX) has revealed the presence of compact
clouds seen in absorption against bright mid-infrared emission from
the Galactic plane (Egan et al. 1998, ApJ 494, L199). Typical column
densities of these dark clouds are estimated to be
N(H$_2$) = 10$^{23-25}$ cm$^{-2}$. About 2000 such clouds were found in a
1\deg\ $\times$ 180\deg\ scan along the Galactic equator.
There is no reason to believe that such clouds would not exist in
other galaxies, even at high redshift.
Any medium to bright GRB happening behind one such  cloud,
seen from redshift 1--3, would be an easy target to detect
nuclear resonance absorption, and allow to study the surroundings
of GRBs and the metallicity of those clouds at large redshifts.

\bigskip\bigskip\bigskip

\noindent
{\Large\bf III. Active Galactic Nuclei}

\vspace{0.3cm}

\begin{figure}[hb]
\includegraphics[bb=77 366 558 692,width=7.5cm,clip]{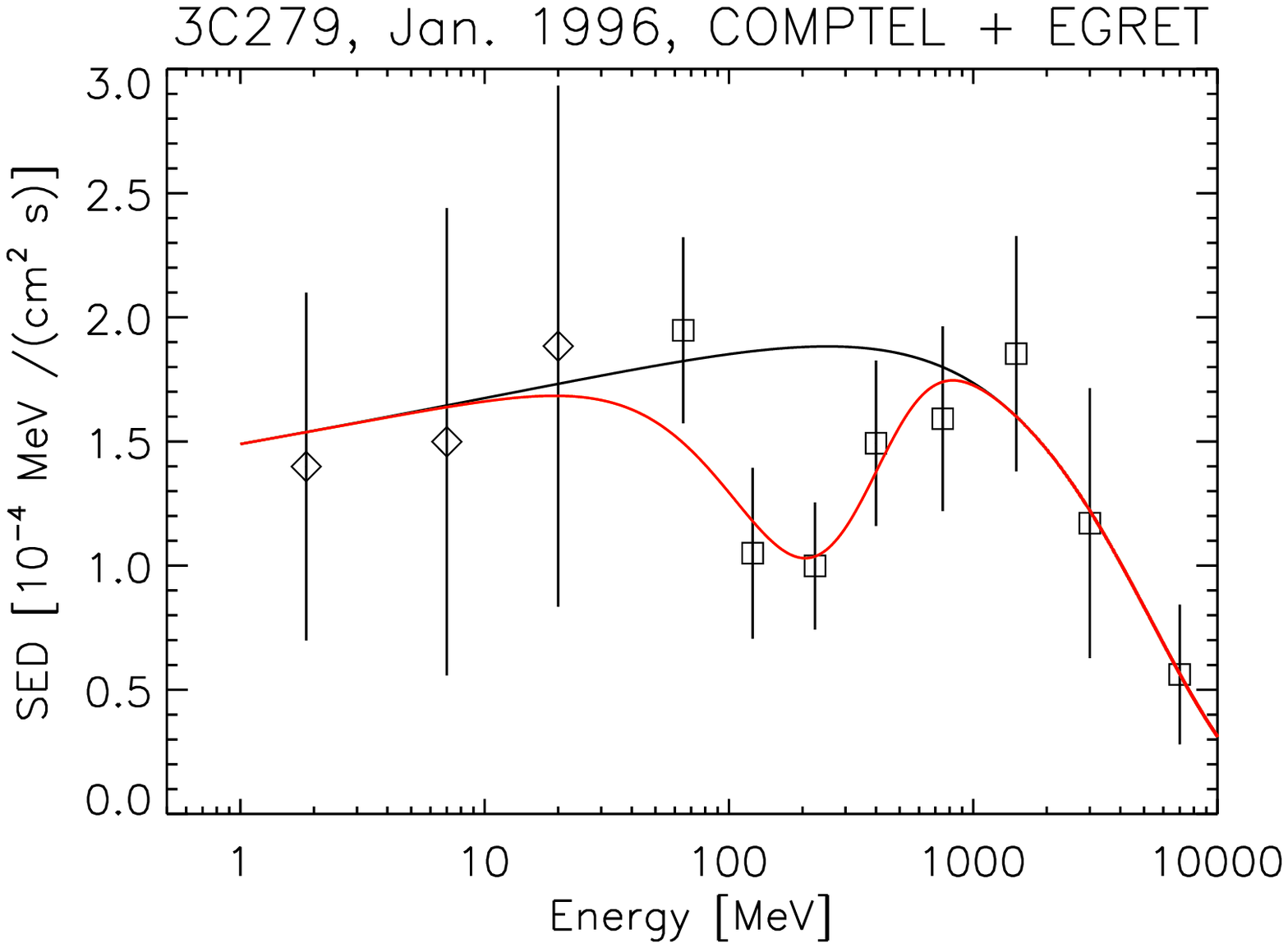}
\includegraphics[width=7.5cm]{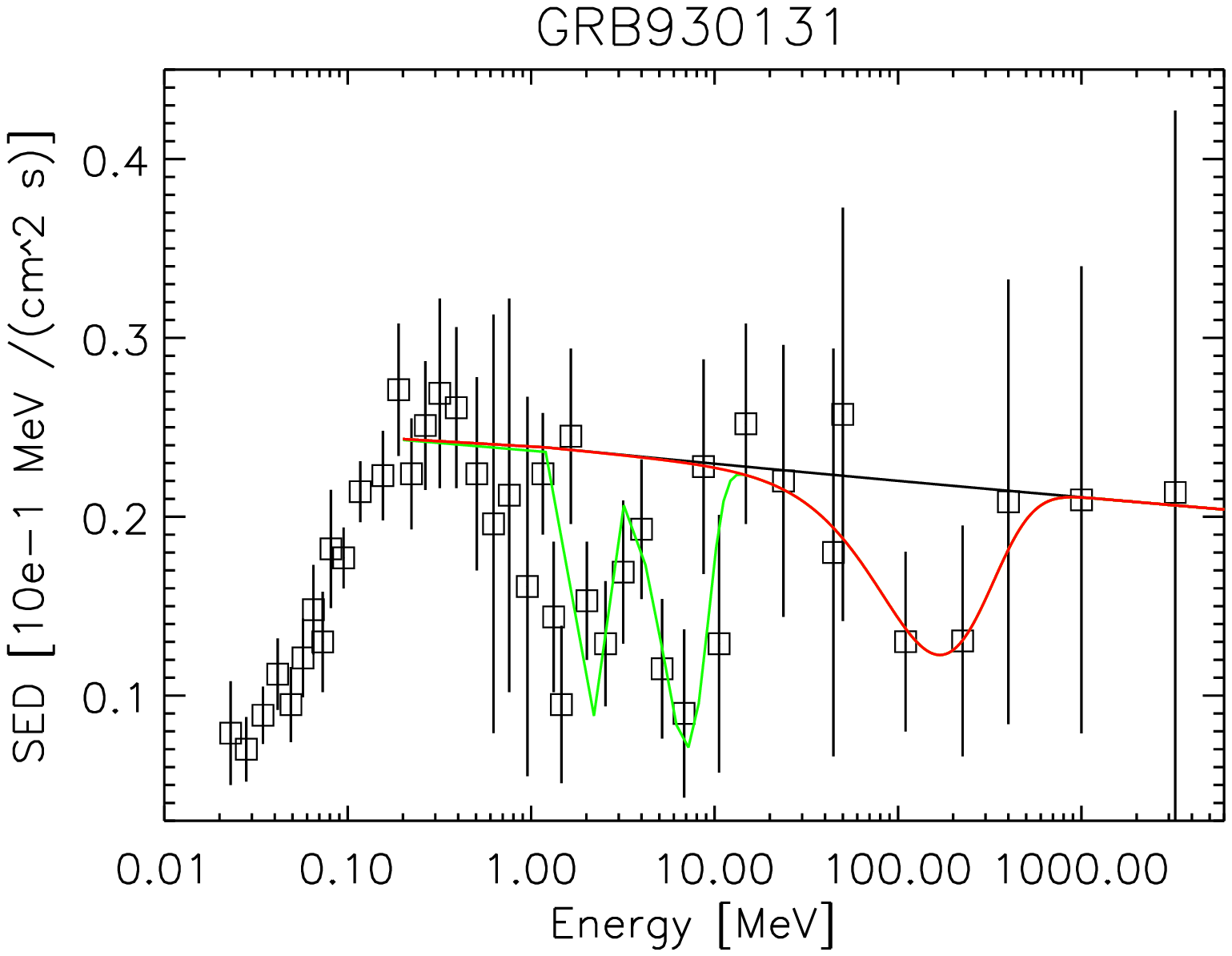}
 \caption[Delta resonance in 3C 279 and GRB 930131]{{\bf Left:}
3C 279 spectrum during the January
1996 flare as measured by COMPTEL (diamonds) and EGRET (squares),
which includes the ${\Delta}$ resonance
absorption in the circumnuclear environment (red line) in addition to
a cut-off power law (black line).
The best-fit energy for the ${\Delta}$ resonance is 208$\pm$25 MeV,
implying a redshift 0.57 $\pm$ 0.12, close to the
optically determined z=0.536.
{\bf Right:}
Fit to the combined COMPTEL/EGRET spectrum of GRB 930131. The throughs at 
5--8 MeV and 100--200 MeV are
compatible with the Giant Dipole and Delta resonance, respectively,
 at  z$\sim$1 (Iyudin \etal\ 2007, in
"Gamma-Ray Bursts: Prospects for GLAST", AIP Conf. 906, p. 89).
The inferred column densities in both cases are 
$\approx$2$\times$10$^{26}$ cm$^{-2}$.
\label{grb93}}
\end{figure}

The class of active galactic nuclei (AGN) includes those high-energy sources
for which the largest column densities have been found so far by
maesurements at hard X-rays or infrared wavelengths. Thus, it is not
surprising that these were also the first class of objects for which
signatures of nuclear resonance absorption has been searched for.
Indeed, the combined spectra of COMPTEL and EGRET, both onboard
the former Compton Gamma-Ray Observatory, have revealed features
in the brightest sources that are at the correct energies.
Despite individual features being at the 1--2$\sigma$ level, 
since these features are seen in different sources located
at different redshifts, and the absorption troughs are seen consistently
at the rest-frame energies of the Delta resonance, the combined
evidence for the reality of these resonance absorptions is remarkable
(Iyudin \etal\ 2005, A\&A 436, 763). Fig. \ref{grb93} shows one example
for 3C 279 as measured during the 1996 flare: when fitting a
Gaussian to the absorption trough, the derived redshift is 0.57,
compared to the optically known redshift of 0.54!

\bigskip

In fact, the potential of measuring redshifts once absorption troughs
are established and are seen in more than a few sources will provide
a new, powerful tool in the identification of new $\gamma$-ray sources
(Iyudin \etal\ 2007, A\&A 468, 919).
EGRET alone left us with about 150 unidentified sources -- 
Fermi/LAT is expected
to provide many more new sources. Measuring the redshifts of these
bright (for LAT standards) EGRET sources would help dramatically
in establishing counterparts at other wavelengths, which has been
a daunting task over the last decade.

Another aspect involves Compton-thick sources. 
The shape of the spectrum of the cosmic X-ray background
(Ajello et al. 2008, ApJ 689, 666) cannot constrain the number of
objects with a column density of $>$10$^{25}$ cm$^{-2}$, the so-called
extra-thick AGN. However, column densities of  $>$10$^{25}$ cm$^{-2}$
have been inferred from the spectra of several reflection-dominated AGN.
Only a method able to measure such column densities in a direct way will
be able to move us beyond the present state of hypotheses.

Unfortunately, the number of objects having spectral energy distributions
with indications of absorption
features remains small due to the low sensitivity
of the previous gamma-ray telescopes in this energy range. Fermi/LAT is
expected to provide new impetus in this direction, though the Delta
resonance is close to the lower energy boundary of LAT, and the
low-energy upturn of the continuum spectrum may be difficult to
establish.

Looking ahead on a time scale of 10 years, nucleonic absorption line 
spectroscopy can be expected to be a growth industry. A mere factor 5
more in instrumental sensitivity over that of presently proposed
instruments will bring column densities of 1$\times$10$^{24}$ cm${-2}$  
in their sensitivity range. These column densities have been seen
already in hard X-ray spectra with Swift/BAT and INTEGRAL/IBIS in AGN.
But many more object classes are then expected to be in reach for
nucleonic absorption line spectroscopy, such as 
inner accretion disks in black hole systems, 
sources buried in dense molecular clouds,
or population III stars if they are powered by dark matter heating 
rather than by fusion (e.g. Freese et al. 2008, 8th UCLA Symp: Sources and 
Detection of Dark Matter and Dark Energy in the Universe, arXiv:0812.4844).

\bigskip\bigskip\bigskip

\noindent
{\Large\bf IV. Fermi/LAT prospects and requirements for a new mission}

\vspace{0.3cm}

Fermi/LAT, in operation since June 2008 and just in the course of
performing a sensitive all-sky survey, is expected to provide the
first proof of existence of nuclear absorption lines in astrophysical
sources, and in turn the existence of source environments with
column densities larger than presently known from INTEGRAL and
Swift/BAT 20--50 keV spectra. In particular, it is the Delta
resonance line at 327 MeV which is in the LAT energy range.
Thus, bright low-redshift sources will imprint a clear signal in
the LAT spectra, and consequently low-redshift AGN like 3C 279
are the most promising sources to discover nuclear absorption lines.
With its all-sky survey, LAT will probe the low-redshift (z$<$1) Universe
for large column densities.
In contrast, for GRBs with typical redshifts beyond 1, and
fluxes above 100 MeV not very high in general (GRB 080916C was
the only exception so far), the prospects provided by LAT are
somewhat worse. Thus, GRBs would benefit from measurements in the lower
MeV range.

For a future mission, this new strategy of using nuclear resonance
absorption requires sensitive spectroscopy
in the 0.5--100 MeV band. The detection of GRBs or AGN flares,
both highly variable objects, requires a large
field of view. Therefore, the logical detection principle is the
combination of Compton interaction and pair creation.
Several detailed mission proposals have been developed over the
last years around the world (one example, which explicitely has
included the quest for nuclear resonance absorption, is described
at http://www.springerlink.com/content/a7148g437188rl44/).

\end{document}